\documentclass[preprint,prmaterials,aps,superscriptaddress,showkeys]{revtex4-2}
\pdfoutput=1

\usepackage[T1]{fontenc}       
\usepackage{xcolor}
\usepackage{booktabs}          
\usepackage{longtable}
\usepackage{graphicx}
\usepackage{tabularx}
\usepackage{hyperref}

\hypersetup{
  pdftitle={By how much can closed-loop frameworks accelerate computational materials discovery?}
  pdfauthor={Vinay I. Hegde <vhegde@citrine.io>}
  colorlinks=true,
  linkcolor=blue,
  citecolor=blue,
  urlcolor=blue,
}
\begin{document}
\title{By how much can closed-loop frameworks\\ accelerate computational materials discovery?}

\author{Lance Kavalsky}\thanks{These authors contributed equally to this work}
\affiliation{Carnegie Mellon University, Pittsburgh, PA 15213}
\author{Vinay I.\ Hegde}\thanks{These authors contributed equally to this work}
\affiliation{Citrine Informatics, Redwood City, CA 94063}
\author{Eric Muckley}
\affiliation{Citrine Informatics, Redwood City, CA 94063}
\author{Matthew S. Johnson}
\affiliation{Massachusetts Institute of Technology, Cambridge, MA 02139}
\author{Bryce Meredig}
\email{bryce@citrine.io}
\affiliation{Citrine Informatics, Redwood City, CA 94063}
\author{Venkatasubramanian Viswanathan}
\email{venkvis@cmu.edu}
\affiliation{Carnegie Mellon University, Pittsburgh, PA 15213}


\begin{abstract}
The implementation of automation and machine learning surrogatization within
closed-loop computational workflows is an increasingly popular approach to
accelerate materials discovery.
However, the scale of the speedup associated with this paradigm shift from
traditional manual approaches remains an open question.
In this work, we rigorously quantify the acceleration from each
of the components within a closed-loop framework for material hypothesis evaluation
by identifying four distinct sources of speedup: (1) task automation, (2)
calculation runtime improvements, (3) sequential learning-driven design space
search, and (4) surrogatization of expensive simulations with machine
learning models.
This is done using a time-keeping ledger to record runs of automated software
and corresponding manual computational experiments within the context of
electrocatalysis.
From a combination of the first three sources of acceleration, we estimate that
overall hypothesis evaluation time can be reduced by over 90\%, i.e., achieving
a speedup of $\sim$$10\times$.
Further, by introducing surrogatization into the loop, we estimate that the
design time can be reduced by over 95\%, i.e., achieving a speedup of
$\sim$$15$--$20\times$.
Our findings present a clear value proposition for utilizing closed-loop
approaches for accelerating materials discovery.

\end{abstract}

\keywords{automation, closed-loop frameworks, DFT, sequential learning, materials discovery}

\maketitle




\section{Introduction}\label{sec:intro}


The discovery and optimization of materials is a central barrier to developing
and deploying next-generation energy technologies~\cite{mistry2021machine}.
In particular, decarbonizing chemical synthesis through electrochemistry
requires the identification of new and efficient
electrocatalysts~\cite{karthish2017electrification}.
One example of such decarbonization is to substitute the energy-intensive
Haber-Bosch process used to synthesize ammonia by materials that can catalyze
the reaction electrochemically~\cite{Suryanto2019, SunNRR_Review}, at
substantially lower energy costs.
However, finding optimal candidates efficiently remains a challenge due to both
the large size of the feasible candidate space~\cite{Kim2020} and the
computational cost of high-fidelity evaluation of each candidate.
Development of methods to accelerate the candidate evaluation search, even
within a well-defined and bounded design-space, is
crucial to meet approaching climate goals.

These considerations have motivated significant research into new methods for
accelerated materials discovery, both experimentally and
computationally~\cite{Tabor2018, Pollice2021}.
In the context of experimental screening, much research focus has
taken the form of robotic experimentation for applications such as searching
for battery electrolytes~\cite{Dave2020otto, dave2022autonomous}, finding
thermally stable perovskites~\cite{Zhao2021perovskite}, and optimizing battery
charging protocols~\cite{Attia2020}.
These studies tend to employ a combination of robots to automate each
experimental task and a learning agent that recommends the next experiment to
perform based on the outputs of previous experiments, thereby ``closing the
loop''.
However, the trade-off is that autonomous experimental setups are highly
application-specific and do not typically probe the material under realistic
device operating conditions.
Thus, although experimental workflows show much promise, they are, at present,
limited in terms of adaptability and bridging the device
gap~\cite{flores2020onthewaytoauto}.

In contrast, computational workflows promise to address a broad range of
material discovery challenges as they are limited only by the availability of
computational resources and the accuracy of the underlying
methods~\cite{coley2020autonomousreview}.
In general, computational workflows share some similarities with closed-loop
experimental workflows, especially around algorithms and approaches for
iteratively selecting the next set of candidates to evaluate from a large
design space.
A notable difference, however, is that any new tasks or pipelines added to a
computational workflow is limited only by computational requirements and not
by the inventory of raw materials, supply logistics, instrumentation setup,
laboratory space, and other considerations.
This allows for improved modularity in existing closed-loop software frameworks
as well as transferability between varying materials discovery workflows.

The use of an iterative informatics-driven search of the design space has
demonstrated encouraging results in terms of speeding up materials
discovery~\cite{warmuth2003active, seko2014machine, pauwels2014bayesian,
chen2015optimal, ward2016general, kiyohara2016acceleration,
podryabinkin2017active, gopakumar2018multi, yuan2018accelerated,
brandt2017rapid, ling2017high, herbol2018efficient, sendek2018machine,
rohr2020benchmarking, del2020assessing, kusne2020fly, gongora2020bayesian}.
Similarly, informatics-driven closed-loop computational workflows have been
shown to discover promising candidates faster than a random search, for
applications such as catalyzing electrochemical CO$_2$ reduction and hydrogen
evolution~\cite{Tran2018}, finding stable iridium oxide
polymorphs~\cite{Flores2020}, and discovering stable binary and ternary
systems~\cite{Montoya2020}.

While closed-loop computational frameworks with embedded guided design space
search demonstrate a promising approach to accelerate materials discovery,
quantification of their benefits over more traditional approaches remains
challenging.
In particular, the degree to which speedups from various components of a fully
autonomous closed-loop framework combine to accelerate materials discovery
remains unclear.
To our knowledge, a detailed breakdown of such sources of acceleration, along
with relative quantitative estimates of the associated speedups, has not been
previously explored.

\begin{figure}[tbh]
  \centering
  \includegraphics[width=0.95\textwidth]{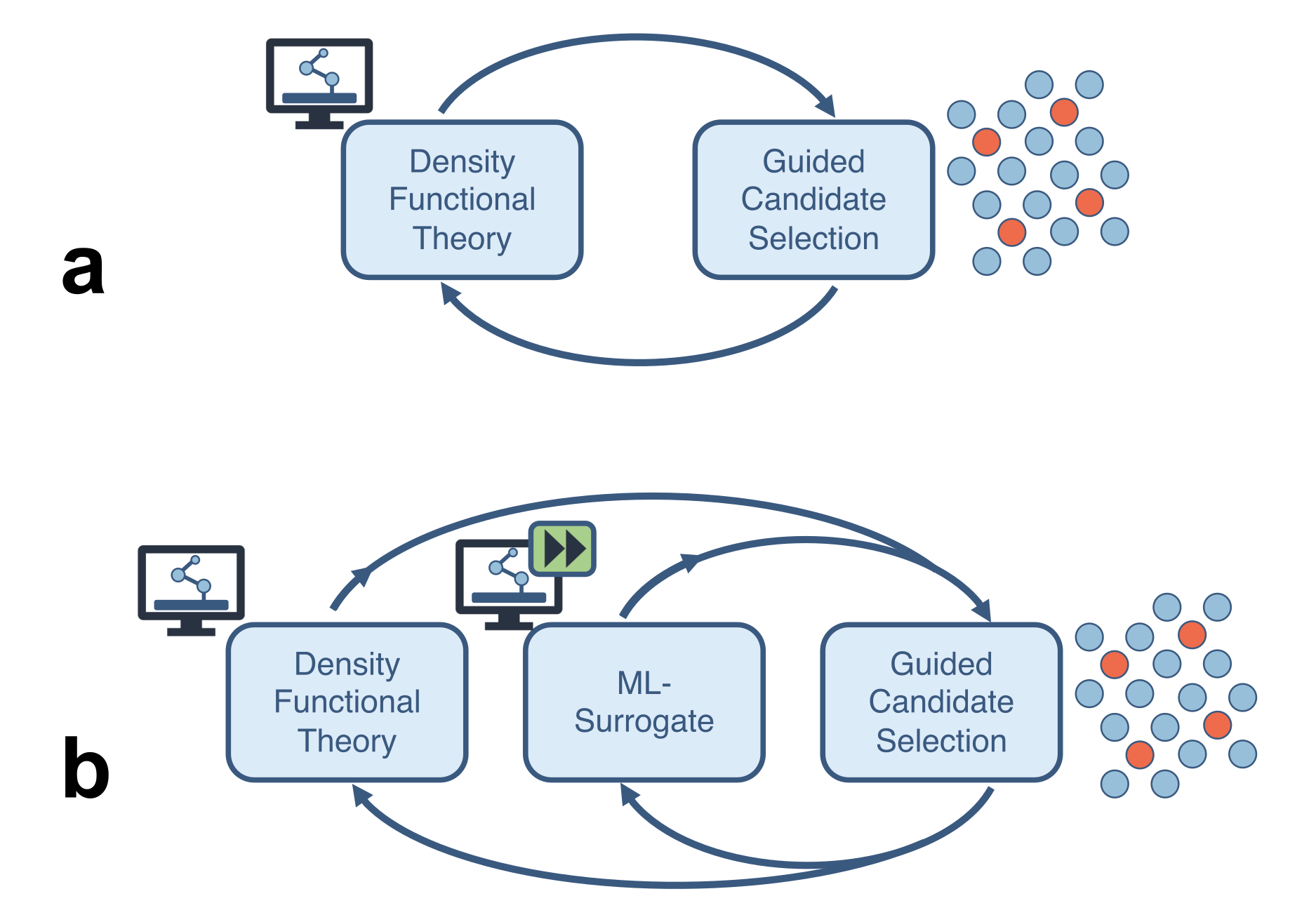}
  \caption{Closed-loop materials discovery frameworks, (a) without, and (b)
  with machine learning surrogates for the density functional theory
  calculations, considered in this work for acceleration
  quantification.}\label{fig:high-level-overview}
\end{figure}

In this study, we quantify the acceleration estimates of a closed-loop
computational framework for an electrocatalysis application.
We probe two types of fully autonomous computational workflows
(Figure~\ref{fig:high-level-overview}):
(a) a closed-loop framework consisting of high-throughput density functional
theory (DFT) calculations which feeds into a sequential learning (SL) algorithm
that selects the next batch of candidate systems (thereby closing the loop), and
(b) an extension of the previous framework where enough DFT data has been
produced to train a machine learning (ML) surrogate to a desired accuracy and
replace the expensive DFT calculations.
We consider four categories of acceleration:
(a) comprehensive end-to-end automation of computational workflows,
(b) runtime improvements of individual compute tasks,
(c) efficient search over vast design spaces using uncertainty-informed SL, and
(d) surrogatization of time-consuming simulation tasks with ML models.

Within each of the above categories, we estimate respective speedups and
aggregate them into overall acceleration metrics.
For end-to-end automation we estimate the attributed speedup through timing
comparisons of automated tasks and their manual analogues.
In addition, we introduce a human-lag model to simulate user-related delays
associated with manual job management on a computational resource.
For runtime improvements, we estimate speedups from using informed calculator
settings as well as better initial structure guesses for DFT structural relaxations.
This comparison is done in the context of calculations for relaxing the OH
moeity onto the hollow sites of a sample single-atom alloy, Ni$_1$/Cu(111).
For efficient design space search, we use a simulated SL-driven process on a
representative problem of finding the bimetallic catalyst with the optimal
surface binding energies for the CO moeity.
For surrogatization with ML models, we estimate the speedup by calculating the
DFT training set size needed to reach a desired model accuracy for adsorption
energy (as opposed to generating the full dataset).
Finally, we accumulate these results into an overall acceleration for workflows
both excluding and including surrogatization.
Through a combination of improvements in each of the above areas, we
demonstrate a reduction in time to discover a new promising electrocatalytic
material by 80-95\% when compared to conventional approaches.

\section{Results}\label{sec:results}
Each of the forms of acceleration described above can synergize to provide
an overall speedup in materials discovery.
We benchmark the acceleration of each individual category through timing
estimates of the relevant components both within a closed-loop automated
workflow and for equivalent tasks when using a more traditional approach.
For the automated workflow, we use the \texttt{AutoCat}, \texttt{dftparse}, and
\texttt{dftinputgen} software packages in tandem.
For the traditional workflow, we record timings for a researcher using the
\texttt{ASE} (Atomic Simulation Environment~\cite{ase-paper}) software package
for equivalent tasks.
Additional details are provided in Section~\ref{sec:methods}.
As an example representative design space, we use the single-atom alloy (SAA)
class of materials.
SAAs are transition-metal hosts whose surface contains dispersed atoms of a
different transition-metal species, and have shown much promise for
electrocatalysis applications~\cite{SykesSAA_review}.

In the following subsections we discuss each of the individual acceleration
categories and how their estimates were obtained.
This is followed by acceleration estimates of the full workflow combining all
sources of speedup to obtain a single acceleration estimate from the automated
closed-loop approach relative to the traditional baseline. 

\subsection{Automation of Computational Tasks and Workflows}

\begin{figure}[tbh]
  \centering
  \includegraphics[width=\textwidth]{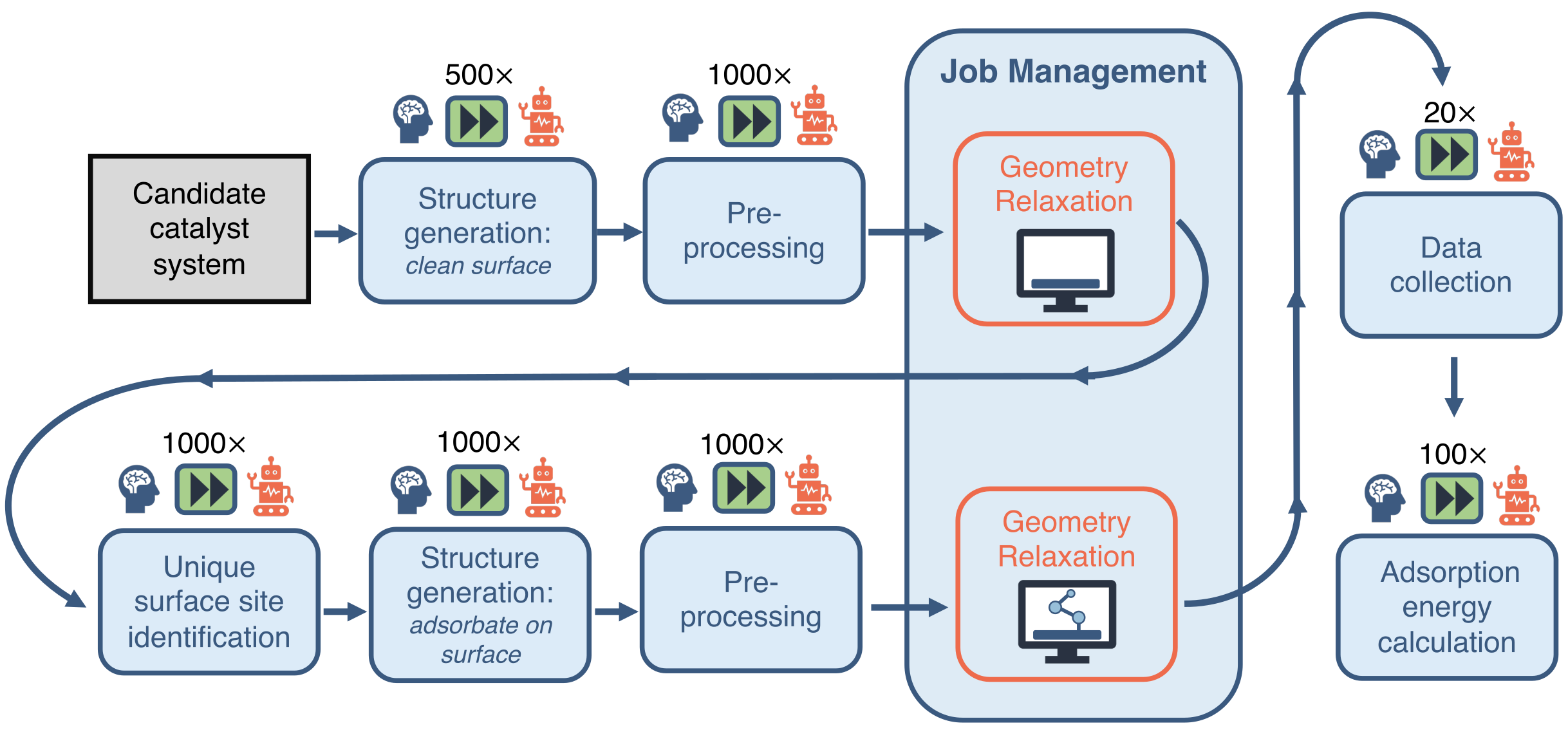}
  \caption{A typical workflow for computational investigation of materials for
  electrocatalysis applications using density functional theory. Blue boxes
  indicate computational tasks which typically require researcher input. Factors
  above each task indicate potential acceleration through automation. Orange
  boxes are geometry optimizations via density functional theory
  calculations.}\label{fig:auto_pipeline}
\end{figure}

Within a standard computational study, there are many time-consuming tasks
related to preparing, managing, and analyzing DFT calculations.
In Figure~\ref{fig:auto_pipeline}, we visualize a typical pipeline for a
computational electrocatalysis study. 
Each of the boxes underneath the symbol of a brain represents a task where user
involvement is required in the traditional paradigm.
This includes structure generation, DFT pre- and post-processing, and job
management on computational resources.
Thus, every box in the pipeline that relies on user intervention is an
opportunity for streamlining through automation.

To benchmark the traditional workflow against an automated one in a fair
manner, we define the same objective for both paradigms: calculation of the
adsorption energies of OH on the SAA of a Ni atom embedded on a Cu (111)
surface, designated as Ni$_1$/Cu(111).
This is further bounded to calculating adsorption only on three-fold sites on
the surface (6 in total). 
The goal is to mimic the scenario where an activity descriptor has already been
identified for a specific electrochemical reaction, thereby collapsing
performance predictions to the surface binding energy of a single adsorbate, as
reported in previous studies~\cite{Tran2018}.
We have recently published methods to identify the most robust descriptors for
a given reaction based on uncertainty quantification
techniques~\cite{Kavalsky2020, Krishnamurthy2018}, and while we focus here on
the binding energy alone, our acceleration estimation methodology is extensible
to more complex descriptors equally well.
As will be discussed later, this task of calculating the surface binding energy
of an adsorbate species is integral to the representative problem of optimizing
the binding energy across a set of possible SAAs using an SL-driven design of
experiments.
It should be noted that while automation generally replaces baseline tasks that
are not very time-consuming in themselves, often on the order of seconds to
minutes, the accelerations reported from this category free up the researcher
to work on more analytical and constructive tasks, as elaborated in
Section~\ref{sec:discussion}.

All of the necessary steps to obtain the specified adsorption energies are
highlighted in Figure~\ref{fig:auto_pipeline}.
A comparison of the estimated time required for each task in the traditional
approach and our automated approach is provided in
Table~\ref{tab:automation-accel}.
Below, we outline the potential acceleration for each task via automation.

\subsubsection{Candidate structure generation}

As an input, DFT requires atomic scale structural representations of the
candidate systems to be evaluated.
Structure generation in the context of electrocatalysis consists of generation
of the catalyst structure without any reaction intermediates, identification of
all of the possible adsorbate sites, and placement of the reaction
intermediates on the sites of interest, along with the potential inclusion of
the effect of water layer~\cite{viswanathan2012simulating}.
In this work, for simplicity, we do not consider solvation effects, but our
analysis framework can be easily extended to include it.
The first task corresponds to writing and executing scripts to
generate the clean Ni$_1$/Cu(111) slab via either \texttt{ASE} or
\texttt{AutoCat} (corresponding to the traditional and automated approaches,
respectively), and recording the relative timings.
While \texttt{ASE} has functions tailored for the generation of some classes of
systems, additional user involvement is necessary for those that are not
currently implemented.
As an example, \texttt{ASE} does not currently have functions geared specifically
towards SAAs, and thus additional scripts are necessary to dope host slabs.
To generate each SAA the dopant site needs to be identified, the substitution
made, and spin polarization added to both the host and dopant, as necessary.
We can contrast this with automation software such as \texttt{AutoCat} which
has a function, built on top of \texttt{ASE} functionalities, to streamline the
generation of these SAA systems.
Further, the implementation in \texttt{AutoCat} is suitable for generating
multiple SAAs through a single function call by the user, including writing the
generated structures to disk in an organized, predictable fashion.
By leveraging tools for streamlined candidate structure generation, a speedup
of approximately $500\times$ over traditional manual approaches is observed.

The estimation of manual site identification for the second task of adsorbate
placement requires measuring the time it takes a graduate researcher to
identify all of the symmetrically unique surface sites of Ni$_1$/Cu(111).
This task becomes increasingly challenging for the researcher as the candidate
catalyst becomes more complex, particularly with broken surface symmetries.
For example, in the case of SAAs, the presence of the single-atom breaks many
of the symmetries, and correctly identifying all unique sites by hand is
nontrivial.
Some sites that are symmetrically equivalent on a non-doped surface no longer
remain so after the substitution of the single-atom. 
In contrast, \texttt{AutoCat} identifies symmetry sites via the Delaunay
Triangulation implementation within the \texttt{pymatgen} software
package~\cite{pymatgen}, providing a systematic automated approach to site
identification that does not require user intervention.
A comparison of the time required for a graduate researcher to identify all of
the sites relative to the automated approach shows a speedup by
a factor of $1000\times$.  

\subsubsection{Density functional theory pre- and post-processing}
For every catalyst structure generated, geometry optimizations via DFT
calculations need to be performed.
The total energies from these relaxed structures can then be used to estimate
properties of interest, such as adsorbate binding energy.
Preparation for each of these calculations involves writing DFT input files and
scripts to submit these calculations to high-performance computing (HPC)
resources.
The DFT input files contain all of the calculation parameters to be used, such
as the k-point density and the exchange-correlation functional.
In addition, job submission scripts contain information about the requested
computational resources on a HPC resource, including the number of compute
cores needed and the walltime at which the job will be forcibly terminated.
To obtain a baseline, we time a user performing both the above tasks, i.e.,
writing scripts to generate DFT input files as well as for submitting batch
jobs to HPC resources.
This is then compared to the time required for the equivalent tasks within a
fully-automated framework using software such as \texttt{dftinputgen} (see
Section~\ref{sec:methods} for details).
The automated tasks are approximately $1000\times$ faster than their
traditional counterparts.

Additionally, once the DFT calculations have successfully completed, the
compilation of results and data can consume a significant amount of time.
The user must read through each of the DFT output files, extract the desired
information, and collect and organize this data.
When scaled up to a large number of systems, and thus calculation outputs, this
can quickly become time-consuming.
Here, we record the time taken to manually read all of the output files and
collect all of the data into a single spreadsheet as well as for the parsing
done by an automated framework using the \texttt{dftparse} and
\texttt{dfttopif} packages (see Section~\ref{sec:methods} for details).
A comparison of the recorded timings shows a speedup for data parsing and
compilation step to be $20\times$.

Once the total energies of the reference states (the SAA surface with/without
the adsorbate and the isolated adsorbate moeity) are extracted, the adsorbate
binding energy can be calculated.
We thus compare the time required to calculate these binding energies within
a spreadsheet manually to that of automatically calculated via a software
framework, resulting in a speedup of $100\times$.
This final post-processing step of calculating the adsorbate binding energies
is relatively quick regardless of the approach taken compared to the other
steps considered in this workflow.

Note that while the speedups from the automation of tasks as described in the
previous two sections are enormous, the baseline estimates for manual
completion of these tasks are quite small, on the order of minutes.
We reiterate that the impact of automation of these tasks is primarily on
researcher productivity, allowing focus on the more analytical tasks rather
than the more routine ones (see Section~\ref{sec:discussion}).

\subsubsection{Workflow integration}
In addition to the automation of candidate structure generation and DFT
pre-/post-processing as described above, the automation of the submission of
batch jobs to HPC clusters, status monitoring, and general job management
provides opportunities for significant acceleration.
DFT calculations of catalyst structures are computationally expensive and
typically require active monitoring by a researcher.
In particular, as these calculations can take variable lengths of time to
complete, they may demand user intervention at unpredictable times.
For example, this could be to fix errors or simply resubmit continuation jobs.
The unpredictability associated with job management introduces ``human lag'' as
it is not possible for the typical researcher to continuously monitor the
status of all submitted DFT jobs at all times.
Here, we estimate such a human lag via a simple Monte Carlo sampling approach.
First, we divide days into three different windows representing typical
working hours, hours where some monitoring may occur, and hours where usually
no monitoring occurs, with ``checkpoints'' in time defined for each (see
Section~I in the SI for details).
Next, we assume a uniform distribution for the job finishing on any day of the
week, without any preference for weekdays or weekends.
This assumption accounts for the fact that often a researcher has no control
over the job queue/priority systems on HPC resources, and a specific
already-submitted job may start whenever resources become available.
Finally, we simulate the process of completion of a DFT job followed by
research action at the nearest checkpoint in time, gathering statistics for
a total of 10$^6$ DFT jobs.
In contrast, since job management within the fully-automated workflow is
handled by a framework involving software such as
\texttt{fireworks}~\cite{fireworks:jain-2015}, there is no equivalent human
lag, which enables significant acceleration.

\begin{table}[tbh]
  \centering
  \begin{tabular}{lrrr}
  \toprule
  Workflow step & Traditional & Automated & Acceleration \\
  \midrule
  \multicolumn{4}{c}{Catalyst structure generation} \\
  \midrule
  Clean surface & 16 min & 2 s & $\sim$500x \\
  Site identification & 10 min &  1 s & $\sim$1000x \\
  Adsorbate placement & 9 min & 1 s & $\sim$1000x \\
  \midrule
  \multicolumn{4}{c}{DFT pre- and post-processing} \\
  \midrule
  Generating DFT input and job management scripts & 9 min & 1 s & $\sim$1000x \\
  Data collection & 3 min & 9 s & $\sim$20x \\
  Adsorption energy calculation & 2 min & 1 s & $\sim$100x \\
  \midrule
  \multicolumn{4}{c}{DFT job submission and management} \\
  \midrule
  Job resubmission and error handling & 9 hr & -- & -- \\
  \bottomrule
\end{tabular}

  \caption{Acceleration from automation of computational tasks and
  workflows.}\label{tab:automation-accel}
\end{table}

\subsection{Calculation Runtime Improvements}\label{ssec:calc_runtime}

In the next category of acceleration, we quantify the speedup of calculation runtimes.
Within our electrochemical materials discovery workflow, the primary physics-based
simulation is DFT.
As these calculations can be time-intensive, improving their runtimes is crucial in
achieving significant acceleration.

In the case of adsorption structures, the initial guesses of the adsorbate geometry can
play a key role.
If the initial guess is far from the ground state geometry, more optimization steps will
be required to reach equilibrium.
Since each step requires a full self-consistent evaluation to obtain the energy and
forces, the initial guess should ideally be as close to the equilibrium as possible to
decrease the overall calculation runtime.
The total runtimes of geometry optimizations via DFT can also be heavily influenced by the
choice of calculator settings, such as initial magnetic moment.
A poor guess of the initial magnetic moment can require more steps to achieve
self-consistency and to converge on the final relaxed value of the magnetic moment.

To decouple the influences of the initial geometry guess and the choice of calculator
settings, we run four sets of relaxations for OH on all of the hollow sites of
Ni$_1$/Cu(111).
We use two initial geometry guesses:
(a) a (chemically) ``informed'' configuration, in which the initial height of the
adsorbate on the catalyst surface is guessed based upon the covalent radii of the
nearest neighbors of the anchoring O atom, and
(b) a ``naive'' configuration, in which the initial height of the adsorbate
is set to 1.5~{\AA} above the catalyst surface, and the OH bond angle is 45$^{\circ}$
with respect to the surface.

In addition to the different initial geometry guesses, we explore two choices for
calculator settings, focusing here on the initial magnetic moment parameter:
(a) a ``tailored'' setting, based on the ground-state magnetic moment of the
single-atom dopant species from the \texttt{ASE} package (thus tailoring the initial
guess for the magnetic moment to the specific SAA system being calculated), and
(b) a ``naive'' setting, using an initial magnetic moment of 5.0~$\mu_{\rm B}$ for the
dopant atom in the SAA, regardless of its identity.
In the specific case of Ni$_1$/Cu(111), since the structure prefers to be in a
spin-paired state (i.e., without spin-polarization), the former approach provides an
initial guess that is closer to the actual spin-polarization of the system.
Note that our intention here is to highlight the impact of these choices on the
acceleration of a DFT calculation, and the choices themselves can originate from
deterministic algorithms, an ML model, or another approach entirely.

\begin{figure}[tbh]
  \centering
  \includegraphics[width=0.97\textwidth]{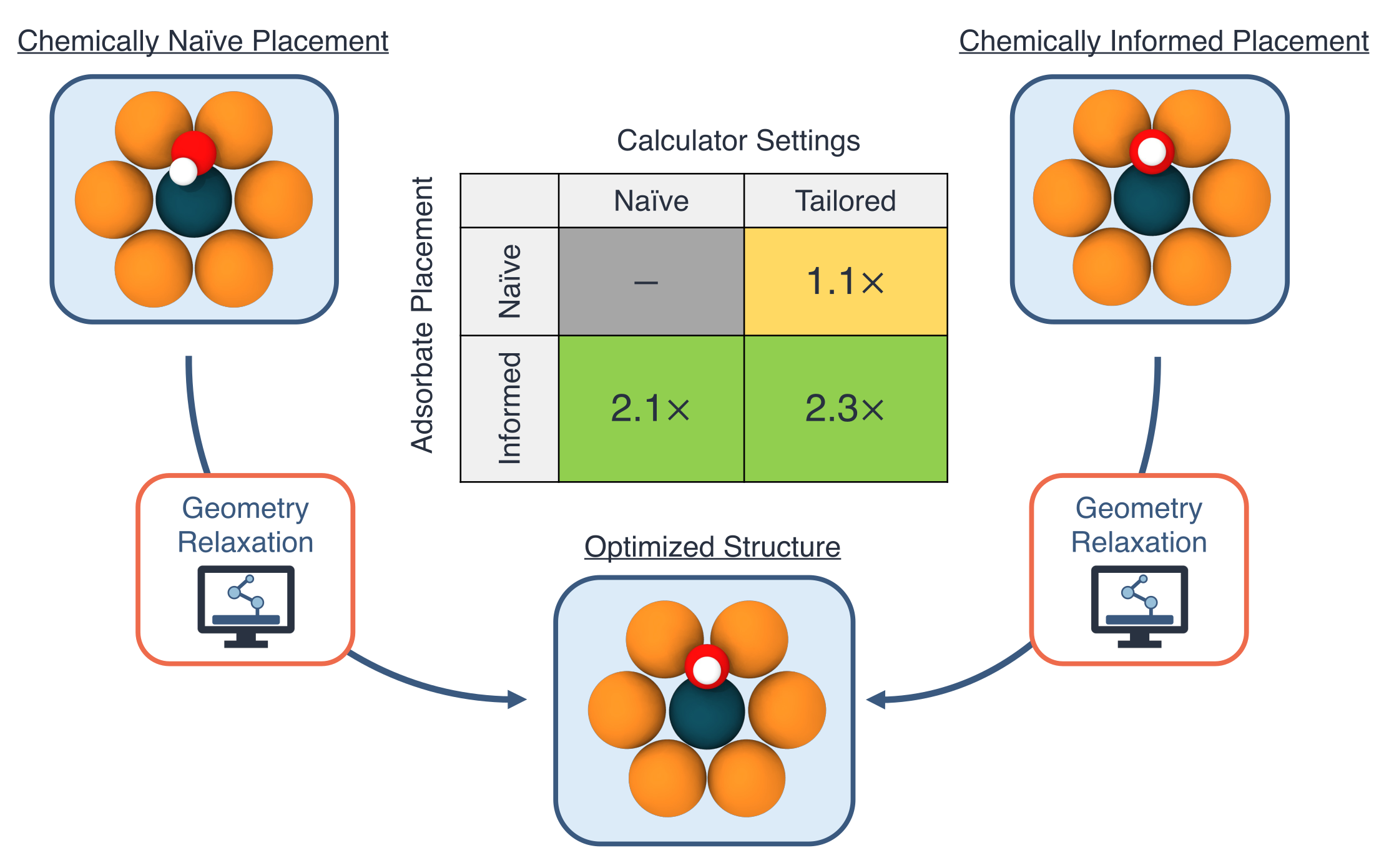}
  \caption{Estimated accelerations for density functional theory geometry optimization
  calculations. The effect of an initial geometry guess and choice of calculator
  settings are decoupled using four independent combinations of informed/naive initial
  geometry and tailored/naive settings. The largest factor of acceleration is observed
  when using an informed geometry guess with a tailored calculator
  settings.}\label{fig:dft_runtime}
\end{figure}

In Figure~\ref{fig:dft_runtime} we visualize the accelerations of the DFT runtimes from
both the choice of calculator settings as well as initial geometry guesses.
Firstly, we observe relatively modest speedups from choice of calculator settings,
approximately $1.1\times$ for both the naive and informed geometry guesses.
For these calculations, the system converges to the non-spin-polarized state within the
first few iterations.
Thus, the observed speedup from the choice of initial magnetic moment of the dopant atom
is mainly a reflection of these initial iterations when the system reaches the
appropriate spin state, which often also take the largest number of self-consistent
steps.
On the other hand, we observe a much larger acceleration from the initial geometry
guess: a speedup of $2.1\times$ and $2.3\times$, for the naive and tailored settings
respectively.

The speedup from a good guess for the initial adsorbate geometry is mainly due to a
reduction in the number of steps required to reach the equilibrium configuration within
a fixed optimization scheme.
For example, an average of approximately 33 and 16 geometry optimization steps using the
Broyden-Fletcher-Goldfarb-Shanno (BFGS) algorithm are required to reach equilibrium,
when starting from the chemically-naive and informed geometries, respectively (with the
tailored calculator settings).
Thus, methods to reduce the number of steps required to reach equilibrium as well as
shorten the DFT compute time at each geometry step (i.e., fewer steps to reach
self-consistency) are highly desirable, and are an area of active
research~\cite{UlissiDOGSS, Boes2019,BligaardBondMin, Deshpande2020, finetuna, gpmin}.
Overall, combining both the improved initial geometry guess as well as the choice of
calculator settings yields the largest factor of runtime acceleration, $2.3\times$, thus
motivating the consideration of both variables within automated workflows.

\begin{table}[tbh]
  \centering
  \begin{tabular}{lrrr}
  \toprule
  Workflow step & Traditional & Automated & Acceleration \\
  \midrule
  \multicolumn{4}{c}{DFT calculation settings and initial structure guess} \\
  \midrule
  Clean substrate relaxation & 21 hr & 18.5 hr & $\sim$1.1x \\
  Substrate + adsorbate relaxation & 46 hr & 20 hr & $\sim$2.3x \\
  \bottomrule
\end{tabular}

  \caption{Acceleration from calculation runtime
  improvements.}\label{tab:runtime-accel}
\end{table}

\subsection{Efficient Design Space Search}\label{sssec:design-space-search}

\begin{figure}[!htb]
  \centering
  \includegraphics[width=\textwidth]{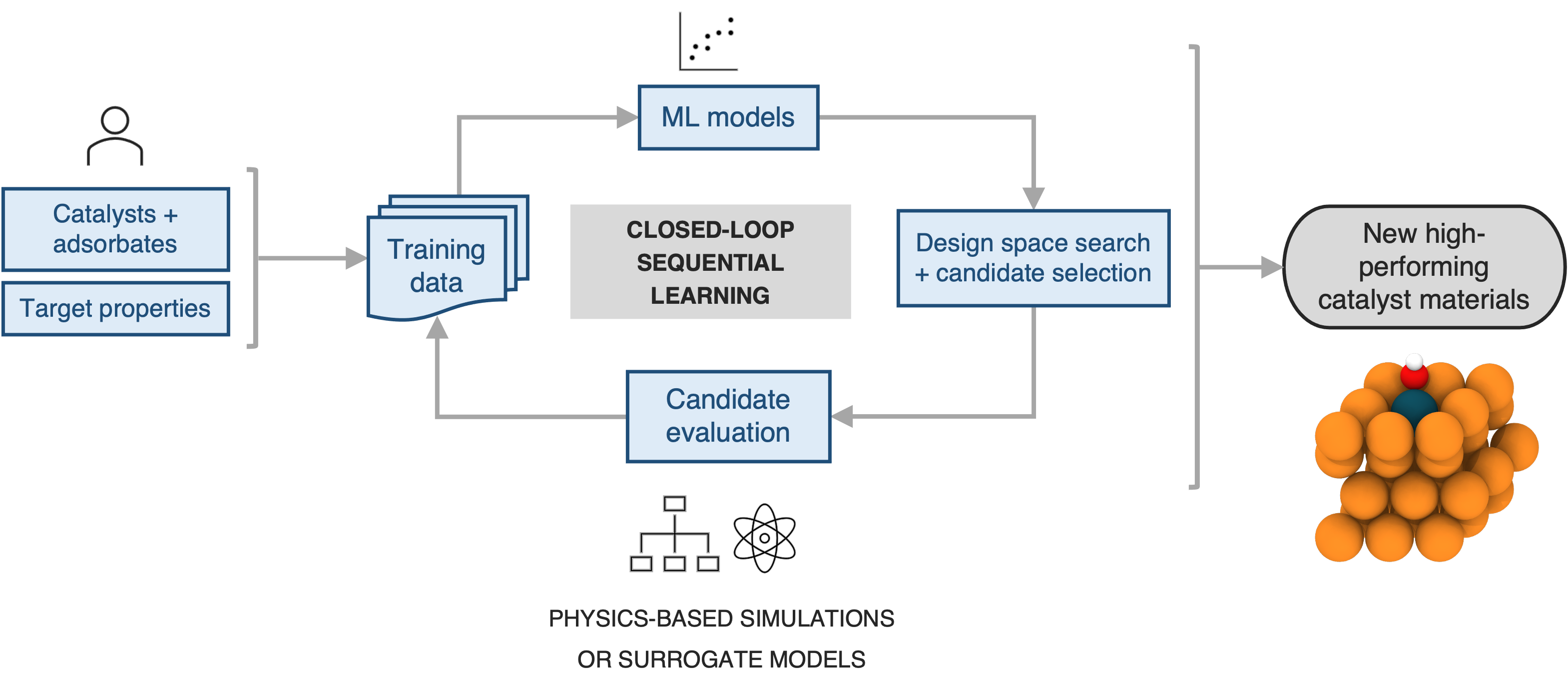}
  \caption{A typical closed-loop sequential learning workflow for computational
  discovery of novel catalyst materials.}
  \label{fig:sl-schematic}
\end{figure}

Next, we estimate the acceleration resulting from use of a sequential learning (SL)
workflow for selecting and evaluating candidates in a design space of catalysts and
compare it to that of traditional approaches.
The SL workflow proceeds as follows:
(1) collect an initial set of a small number of training examples of catalyst candidates
and their properties;
(2) build ML models using the initial set of training examples and predict the objective
properties of all the candidates in the design space of interest;
(3) use an acquisition function that considers model predictions \textit{and}
uncertainties to select the next candidate to evaluate;
(4) evaluate the selected candidate and add it, along with its newly obtained property
values, to the training set;
(5) iterate steps 2--4 in a closed-loop manner until a candidate, or a certain number of
candidates, with the target properties has been discovered.
A detailed schematic of this workflow is presented in Figure~\ref{fig:sl-schematic}.
Such a strategy has been previously shown to be more efficient in sampling the design
space to find novel candidates by a factor of 2--6$\times$ over traditional grid-based
searches or random selection of candidates from the design
space~\cite{warmuth2003active, seko2014machine, pauwels2014bayesian, chen2015optimal,
ward2016general, kiyohara2016acceleration, podryabinkin2017active, gopakumar2018multi,
yuan2018accelerated, brandt2017rapid, ling2017high, herbol2018efficient,
sendek2018machine, rohr2020benchmarking, del2020assessing, kusne2020fly,
gongora2020bayesian}.

\begin{figure}[tbh]
  \centering
  \includegraphics[width=\textwidth]{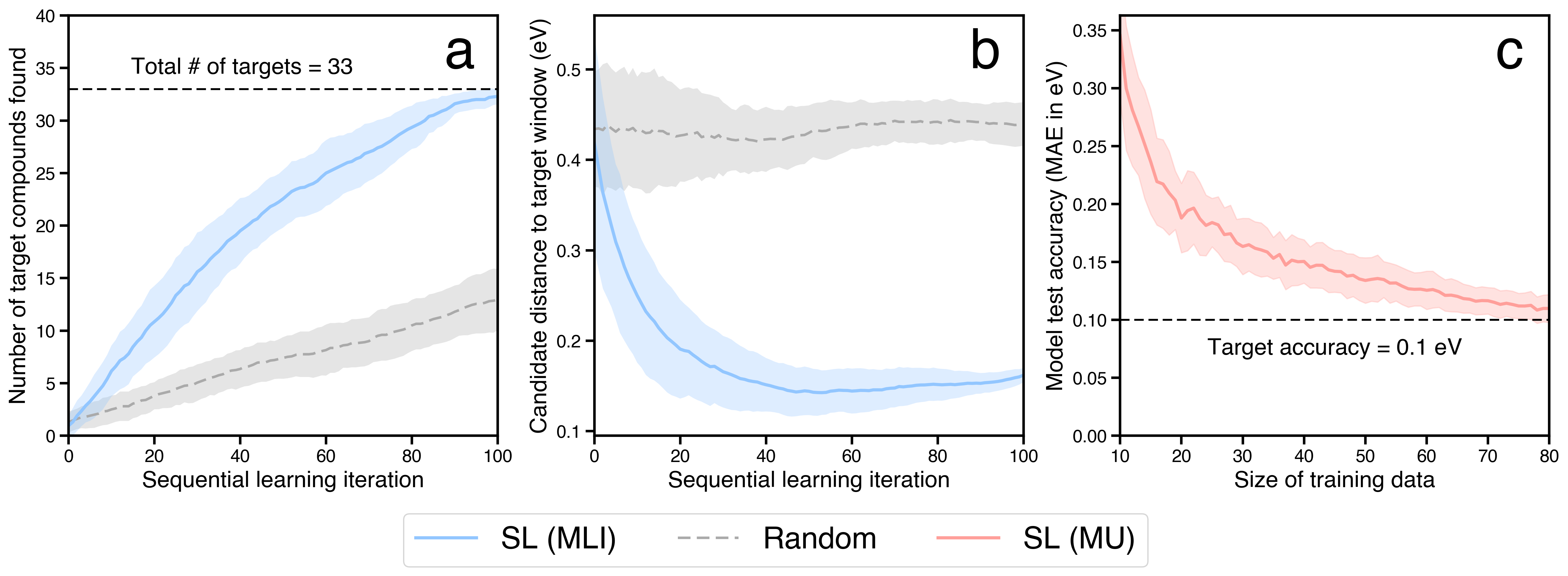}
  \caption{
    A comparison of random search vs sequential learning (SL)-driven approach to find
    new bimetallic catalysts with a target property. (a) Overall, the SL-driven approach
    identifies all the 33 target candidates in the dataset within 100 iterations,
    $\sim$3$\times$ faster than random search. (b) Candidates surfaced via SL lie much
    closer to the target window on average, when compared to those chosen via random
    search. (c) An SL-driven approach can help identify a much smaller number of
    examples that can be used to train ML surrogates to a desired accuracy, at a
    fraction of the overall dataset size. Here, the overall dataset has $\sim$300
    candidates, and an ML model trained on only $\sim$25\% of the candidates chosen via
    a SL-driven maximum uncertainty-based approach achieves the target accuracy.
  }\label{fig:design-space-search}
\end{figure}

For benchmarking the acceleration from SL for a typical catalyst discovery problem, we
use a dataset of $\sim$300 bimetallic catalysts for CO$_2$
reduction~\cite{ma2015machine}.
The dataset contains $\sim$30 candidates with the target property of $^*$CO
adsorption energy on the catalyst surface inside a narrow window of [$-0.7$~eV,
$-0.5$~eV].
We perform an SL simulation, starting with a small initial training set of 10 examples
from the above dataset, and iterate in a closed-loop as described above until all the
target candidates in the design space have been identified successfully, and benchmark
the acceleration against random search.
In particular, at each SL iteration, we build random forest-based models using the
\texttt{lolo} software package~\cite{ling2017high}, and predict the $^*$CO adsorption
energies of all candidates, along with robust estimates of uncertainty in each
prediction.
The next candidate to evaluate is chosen based on the maximum likelihood of improvement
(MLI) acquisition function. This function selects the system with the maximum likelihood
of having an adsorption energy in the [$-0.7$~eV, $-0.5$~eV] window, when considering
both the predicted value as well as its uncertainty.
Overall we find that such an SL-based workflow successfully identifies all $\sim$30
target candidates 3$\times$ faster than random search (Figure~\ref{fig:design-space-search}a).
In addition, we note that the candidates surfaced by SL, on average, have properties
closer to the target property window than those surfaced by random search, even when
those candidates do not explicitly fall within the window
(Figure~\ref{fig:design-space-search}b).
In other words, in addition to discovering target candidates considerably more
efficiently than random search, an SL-based approach surfaces potentially interesting
candidates near the target window much more frequently as well.

\subsection{Surrogatization of Compute-Intensive Simulations}

For the last category of acceleration, we estimate the extent of further possible
speedup through the surrogatization of the most time-consuming tasks in the workflow.
In particular, the rate determining step of the closed-loop framework considered here is
the calculation of the binding energies of adsorbates using DFT. 
ML models can be used as surrogates for physics-based simulations of material
properties often at a fraction of the compute cost and with marginal loss in accuracy.
The primary cost of building such ML surrogates for materials properties often lies in
the generation of training data where such data does not exist, especially when the data
generation involves compute-intensive physics-based simulations such as DFT.
Here we estimate the size of such training data required to build and train ML
surrogates with a target accuracy, and in particular, when such training data is
iteratively built using an SL-based strategy.

We use the dataset of bimetallic catalysts for CO$_2$ reduction mentioned in
Section~\ref{sssec:design-space-search} within a SL workflow to simulate an
efficient, targeted training set generation scheme.
Similar to the SL workflow employed in the search for novel catalyst materials in a
design space of interest, we employ a closed-loop iterative approach to generate the
training data and address model accuracy.
We consider a small initial training dataset of 10 systems, build random forest models
to predict adsorption energies, and iteratively choose the next candidate to build the
training data.
With model accuracy in mind, we employ an acquisition strategy that optimizes for the
most accurate ML model \textit{on average} by choosing candidates to evaluate from
regions in the design space where the model is the least informed.
In particular, at each iteration the candidate whose property prediction has the maximum
uncertainty (MU) is selected to augment the training data. 
The inclusion of such a candidate results in the highest improvement in the overall
accuracy of the ML model.
Note that the aim of using the MU acquisition function is to build a minimal dataset
that is nearly as informative as the full dataset.
This is in contrast to the previously described strategy of using the MLI acquisition
function for discovering the top-performing candidates as quickly as possible.
Using an accuracy threshold of interest, we then determine the fraction of the overall
training data necessary for building useful ML surrogates.
For instance, with a threshold of 0.1~eV (the typical difference between DFT and
experimental formation energy values \cite{kirklin2015open}), we estimate that accurate
ML surrogates can be trained using a dataset generated via the above SL-strategy with
$\sim$25\% of the overall dataset size (Figure~\ref{fig:design-space-search}c).
The accuracy metric here is calculated on a test set of fixed size, via a bootstrapping
approach, as described in the Supplementary Information.

\subsection{Overall Acceleration of the Full End-to-end Workflow}\label{ssec:overall-accel}

Finally, we aggregate the acceleration from the various steps in the workflow to
estimate the overall speedup achieved.
Here, we use the single-atom alloys (SAA) design space for calculating the overall
estimates.
We begin by estimating the size of such a design space.
Limiting the design space to $\sim$30 transition metal hosts and dopants
results in a total of $^{30}$C$_2$ $\approx 900$ SAA systems.
For each SAA system, typically a few (3--5) low-index surface terminations are
considered.
Moreover, the considered adsorbate molecule can adsorb onto the catalyst surface at one
of many possible symmetrically unique sites (up to 20--40 configurations), and all such
possible intermediate configurations need to be considered in the design space.
Overall, a typical SAA design space when fully enumerated can have up to 10$^5$--10$^6$
possibilities.

Using the above SAA design space, we apply the estimated time for each step in our
overall end-to-end catalyst workflow as described in the previous sections, using both
traditional and automated closed-loop methods (with and without surrogates), and
calculate the overall speedup.
From the automation of tasks and workflows, and runtime improvements alone, an
acceleration of $\sim$10$\times$ (a reduction of $\sim$90\%) over traditional materials
design workflows can be achieved.
Further utilizing the ML surrogates (including the compute costs required to
generate the training data) can result in an acceleration of up to
$\sim$20$\times$ (a reduction of up to $\sim$95\%) over traditional approaches.

\begin{table}[tbh]
  \centering
  {\scriptsize 

\begin{tabular}{c|c|c|c|c|c|c|c|c}
  \toprule
  Approach & Structure & Substrate & Adsorbate & Catalyst & Data & Post-processing & Design Space & Total \\
   & Generation & Calculation & Placement & Calculation & Usage & & Search Factor  & Acceleration \\
  \midrule
  Traditional & 16 min & 21 hr & 18 min & 72 hr/i.c.$^1$ & 100\% & 5 min & 1 & \\
  Automated & 2 s & 18.5 hr & 2 s & 20 hr/i.c.$^1$ & 100\% & 10 s & 0.33 & 10$\times$ \\
  + Surrogates & 2 s & -- & 2 s & 20 hr/i.c.$^1$ & 10-25\%$^2$ & 2 s & 0.33 & 15-20$\times$ \\
  \bottomrule
\end{tabular}

}
  \caption{
    Overall acceleration benchmarks for the end-to-end workflows, with and without
    surrogatization. We demonstrate a speedup of up to 10$\times$ with automation of
    tasks and runtime improvements, and a speedup of up to 20$\times$ upon using ML
    surrogates for the most compute-intensive DFT tasks. $^1$i.c.\ = intermediate
    configuration (total \# i.c.\ $\approx 200$/catalyst system); ``traditional''
    includes human lag estimates. $^2$estimate from bimetallic catalyst dataset of the
    relative amount of DFT training data needed to reach a target accuracy of
    0.1~eV/adsorbate.
  }\label{tab:overall-accel}
\end{table}

\section{Discussion}\label{sec:discussion}

The results presented here have implications that reach beyond the reported factors of
acceleration.
It is helpful to make a distinction between project time and researcher time.
We consider project time as the time necessary to carry a project to completion.
In other words, this is an accumulation of all the time spent towards achieving the
tasks to reach the project goal.
Thus, all the acceleration factors quantified above are with respect to this project
time.
Therefore, the closed-loop workflows discussed here are anticipated to have a direct
impact on time to project completion.
In addition, by breaking down the acceleration factors for each component of the
workflows, estimates for project time acceleration in the case of differing closed-loop
framework topologies than those outlined here (e.g., a framework with multi-scale
simulations in place of or in addition to DFT calculations) can be inferred.

On the other hand, researcher time can be interpreted as time spent from the frame of
reference of the researcher on a given workday.
The acceleration associated here is not directly quantified as with project time.
The most obvious example of this influence is through task automation.
In the traditional paradigm, these tasks can become time-consuming, particularly as the
scale and throughput of the project increase.
Automation frees up valuable researcher time that would normally be occupied by the more
mundane tasks.
This allows the researcher to instead focus on more intellectually demanding tasks such
as surveying existing literature, refining the design space and project formulation, and
improving research productivity.

The automation of job management has the benefit of impacting both project time and
researcher time.
Since this form of automation facilitates running computational jobs around-the-clock,
the human-lag associated with monitoring and handling jobs manually is entirely removed.
This decreases the project time as described above.
In the context of researcher time, this automation also has the added benefit of
decreasing the overhead of job monitoring at regular intervals.

We can make a few additional observations regarding the nature of the baselines used to
estimate the speed of traditional approaches in this work.
First, for estimation of task timings such as input file generation for simulations and
script generation to submit jobs on HPC resources, we use time estimates from a single
researcher.
The timings of such tasks are inherently variable, depending on the exact nature of the
task, the researcher performing it, as well as the environmental
setup in which it is performed.
Similarly, natural delays associated with monitoring and managing ongoing computational
jobs depend on the working habits of the researcher, the time-scale associated with each
computation (e.g., those that take hours opposed to days or weeks to complete), and the
availability or connectivity of the computational resources (e.g., on-site resources
versus those that can be accessed remotely).
Lastly, to estimate the acceleration from an intelligent exploration of the design space
using sequential learning, we use random sampling as the benchmark.
While random sampling is an excellent exploratory acquisition
function~\cite{Bergstra2012}, it is not a substitute for traditional methods of design
space exploration.
Typically, traditional search approaches are influenced by prior knowledge, research
directions within the community at the time, available resources, among other factors.
We use random search here, not least because a model to predict a traditional materials
design trajectory does not exist, to our knowledge, but also because it is widely-used
as an unbiased exploratory baseline \cite{warmuth2003active, seko2014machine,
pauwels2014bayesian, chen2015optimal, ward2016general, podryabinkin2017active,
gopakumar2018multi, yuan2018accelerated, ling2017high, herbol2018efficient,
sendek2018machine, rohr2020benchmarking, del2020assessing, kusne2020fly,
gongora2020bayesian}.

We want to emphasize that, given some of the variability in the baselines as discussed
above, the goal of this work is to highlight the approximate scale of acceleration that
can be attributed to the several individual components in a closed-loop computational
materials design workflow.
Moreover, we also aim to highlight the challenges associated with estimating such
factors of acceleration, versus attempting to maximize the accuracy of each timing
estimate itself.
Further methodological improvements for more precisely determining accelerations
associated with each step in an automated workflow would be a valuable area for further
study.
Our work underscores the importance of data collection and sharing, especially around
time spent on research tasks, monitoring and managing medium- to high-throughput
computational projects, implementing traditional approaches of materials discovery and
design trajectories, and handling failed computations and experiments.
We recommend a community-driven initiative towards such data collection and sharing
efforts to bolster our understanding of the traditional baselines as well as to further
contextualize the significant benefits of automation and ML-guided strategies.


\section{Conclusion}\label{sec:conclusion}
In this work we demonstrate that task automation and runtime improvements combined with
a sequential learning-driven closed-loop search can accelerate a materials discovery
effort by more than 10$\times$ (or more than 90\% reduction in overall time/cost) over
traditional approaches.
Further, we estimate that such automation frameworks can have a significant impact on
researcher productivity (20--1000$\times$), direct compute costs (1.1--2.3$\times$), and
project/calendar time ($>$10--20$\times$).
Using a comparison of recorded times for manual computational experiments versus
fully-automated equivalents, we provide speedup estimates stemming from different
components within a closed-loop workflow.
The automation of tasks helps in streamlining, minimizing or completely eliminating the
need for user intervention.
We also identify that significant speedup in terms of simulation (here, DFT calculations) runtimes can be achieved through better initial prediction of the
catalyst geometries as well as better choices for calculator settings.
Moreover, the use of a sequential learning framework to guide the design of experiments
can dramatically decrease the number of candidate evaluations required to achieve the
target materials design goal.
Finally, we extend this analysis to include replacement of time-consuming simulations
with machine learning surrogates, another source of acceleration, and find an
improvement in the overall speedup to $>$15--20$\times$ (or more than 95\% reduction in
the overall time/cost).
We believe that our findings underscore the immense benefits of introducing automation,
machine learning, and sequential learning into scientific discovery workflows, and
motivate further widespread adoption of these methods.

\section{Methods}\label{sec:methods}

\subsection{Workflow Topology}

We consider two different closed-loop ``topologies''.
The first is a two-stage process consisting of DFT calculations to calculate adsorption
energies which are then used in a sequential learning (SL) workflow to iteratively guide
candidate selection (Figure~\ref{fig:high-level-overview}a).
Each DFT calculation task, here, for an electrocatalysis problem, consists of multiple
steps.
Namely, a geometry relaxation of the ``clean'' catalyst surface (i.e., without
reaction intermediates), followed by a geometry relaxation of all reaction intermediates adsorbed
onto all symmetrically-unique sites on the (relaxed) catalyst surface.
In an automated workflow, these DFT calculations are performed sequentially within a
predetermined pipeline framework.
Here, we use a combination of \texttt{AutoCat}
(\url{https://github.com/aced-differentiate/auto_cat}) for automated generation of
catalyst and adsorbate structures, and the \texttt{dftinputgen} and \texttt{dftparse}
software for the DFT calculations.
More details on these software packages are provided in
Section~\ref{ssec:auto_software}.

Another topology we consider is an extension of that described above, with machine
learning (ML) models used as surrogates for the DFT calculations
(Figure~\ref{fig:high-level-overview}b).
In this scenario, the first few overall SL iterations proceed the same as before, except
now as the DFT data is generated, a surrogate ML model is trained on the resulting data
until a threshold test accuracy is reached.
For these first few ``data generation'' iterations, candidates are selected with the
intent of improving overall prediction accuracy.
Once the threshold accuracy for the surrogate model is met, all subsequent iterations of
the loop will use the surrogate model only (instead of DFT calculations) to predict
adsorption energies.
From this point onward the candidate selection step in the SL workflow is then focused
on identifying the most promising materials, as described above in the topology without
surrogatization.

\subsection{Automation Software}\label{ssec:auto_software}

To create the crystal structures for the DFT calculations, we use \texttt{AutoCat}, a
software package with tools for structure generation and sequential learning for
electrocatalysis applications.
This package is built on top of the Atomic Simulation Environment
(\texttt{ASE})~\cite{ase-paper} and \texttt{pymatgen}~\cite{pymatgen} to generate the
catalyst structures \textit{en masse}, and write them to disk following an organized
directory structure.
\texttt{AutoCat} has tailored functions for generation of single-atom alloy (SAA)
surface structures, with optional parameters such as supercell dimensions, vacuum
spacing, and number of bottom layers to be fixed during a DFT relaxation, with
appropriate defaults for each parameter.
Moreover, through the use of \texttt{pymatgen}'s implementation of Delaunay
triangulation~\cite{montoya2017delaunay}, the identification of all of the unique
symmetry sites on an arbitrary surface is automated.
Furthermore, initial heights of adsorbates are estimated using the covalent radii of the
anchoring atom within a given adsorbate molecule as well as its nearest neighbors host
atoms on the surface.
As the development of this package is part of an ongoing work, additional details will
be reported in a future publication.

Once the catalyst and adsorbate systems have been generated by \texttt{AutoCat}, the
crystal structures are used as input to an automated DFT pipeline that
(a) generates input files for a DFT calculator (here we use \texttt{GPAW}~\cite{GPAW1,
GPAW2}),
(b) executes DFT calculation workflows, and 
(c) parses successfully completed calculations and extracts useful information.

\noindent
\textit{Automatic DFT input generation:}
We leverage the Python-based \texttt{dftinputgen} package
(\url{https://github.com/CitrineInformatics/dft-input-gen}) to automate the
generation of DFT input files from a specified catalyst/adsorbate crystal
structure.
In particular, we extend the \texttt{dftinputgen} package to support \texttt{GPAW}.
For a given input crystal structure, the package provides sensible defaults to
use for commonly-used DFT parameters based on prior domain knowledge for novice
users as well as fine-grained control over each parameter for more experienced
DFT practitioners.
The package also implements, ``recipes'', sets of DFT parameters and values to
be used as default depending on the properties of interest, e.g., ground-state
geometry and electronic structure.
The package outputs input files in a user-specified location that can be
directly used by popular DFT packages as input for calculation.

\noindent
\textit{Execution of DFT calculation workflows:}
We leverage the Python-based
\texttt{fireworks}~\cite{fireworks:jain-2015} package to both define complex
sequences of DFT calculations necessary for electrocatalysis studies 
(e.g., clean surface relaxation followed by adsorption relaxation), and to
create, submit, and monitor batch jobs on high performance compute (HPC) resources for
each such calculation.
These scripts are part of an ongoing study and will be open-sourced.

\noindent
\textit{Parsing output from DFT:}
After the completion of DFT calculations of a large number of different candidate
systems, key metrics such as total energy and forces need to be extracted.
To accomplish this task we have extended the previously-developed \texttt{dfttopif}
(\url{https://github.com/CitrineInformatics/pif-dft}) and \texttt{dftparse}
(\url{https://github.com/CitrineInformatics/dftparse}) packages to parse output
generated via \texttt{GPAW}.
Functions written for this package can look for a \texttt{.traj} file resulting from a
successful \texttt{GPAW} calculation in a specified directory.
Once a \texttt{.traj} file has been identified, it can be read using \texttt{ASE} to
extract calculated properties of interest.
This includes not only results such as total energy and forces, but also calculator
settings such as the exchange-correlation functional used.
The extracted properties findings are then written into a Physical Information File
(PIF)~\cite{michel2016beyond} (\url{https://citrine.io/pif}), a general-purpose
materials data schema, for every calculation conducted.

\subsection{First-Principles Calculations}
All DFT calculations are performed with the \texttt{GPAW} package~\cite{GPAW1,
GPAW2} via \texttt{ASE}~\cite{ase-paper}.
The projector-augmented wave method is used for the interaction of the valence electrons
with the ion cores.
A target spacing of 0.16~{\AA} is applied for the real-space grid, with a Monkhorst-Pack
\cite{Monkhorst1976SpecialIntegrations} k-mesh of 4$\times$4$\times$1 for all surface
calculations.
For improved self-consistent field convergence, a Fermi-Dirac smearing of 0.05~eV is
applied.
All geometry optimizations are conducted via the BFGS algorithm as implemented in
\texttt{ASE}.



\subsection{Machine Learning Models}\label{ssec:methods_ml}

We use ML models based on random forests \cite{breiman2001random} as described in the
previously-reported FUELS framework~\cite{ling2017high}.
The uncertainty in a model prediction is determined using jackknife-after-bootstrap and
infinitesimal jackknife variance estimators~\cite{wager2014confidence}.
All ML models and related analysis in this work use random forests and uncertainty
estimates as implemented in the open-source \texttt{lolo} library~\cite{lolo-github}.
Materials in the training dataset are transformed into the Magpie
features~\cite{ward2016general}, a set of descriptors generated using only the material
composition, as implemented in the \texttt{matminer} package~\cite{ward2018matminer}.

\section*{Conflicts of interest}\label{sec:coi}
There are no conflicts to declare.

\section*{Acknowledgements}\label{sec:ack}
The work presented here was funded in part by the Advanced Research Projects
Agency-Energy (ARPA-E), U.S. Department of Energy, under Award Number
DE-AR0001211.
L.K. acknowledges the support of the Natural Sciences and Engineering Research
Council of Canada (NSERC).
The authors thank Rachel Kurchin for helpful discussions around automation and acceleration estimation, and James E. Saal for providing comments on a previous version of this manuscript.

\section*{Author Contributions}\label{sec:author_contrib}
Conceptualization: B.M., L.K., V.H., V.V.;
Methodology: B.M., L.K., V.H., V.V.;
Software: E.M., L.K., V.H.;
Validation: L.K., V.H.;
Data Curation: L.K., V.H.;
Writing -- Original Draft: E.M., L.K., M.J., V.H.;
Writing -- Review \& Editing: all authors;
Visualization: L.K., V.H.;
Supervision: B.M., V.V.

\section*{Data Availability}\label{sec:data_avail}
All data and Python scripts required to perform the analysis presented in this
work are made available via the GitHub repository at
\url{https://github.com/aced-differentiate/closed-loop-acceleration-benchmarks}.
Data shared includes data processing and calculation timing records, crystal
structure files, and a preexisting catalysts dataset used for benchmarking.
Scripts shared include those for estimating human lag in job management,
calculating acceleration from sequential learning, performing all related data
aggregration, analysis, and reproduction of associated figures.

\clearpage


\bibliography{references}

\end{document}


\title{Supplementary Information:\\ 
By how much can closed loop frameworks\\ accelerate computational materials discovery?}

\author{Lance Kavalsky}\thanks{These authors contributed equally to this work}
\affiliation{Carnegie Mellon University, Pittsburgh, PA 15213}
\author{Vinay I.\ Hegde}\thanks{These authors contributed equally to this work}
\affiliation{Citrine Informatics, Redwood City, CA 94063}
\author{Eric Muckley}
\affiliation{Citrine Informatics, Redwood City, CA 94063}
\author{Matthew S. Johnson}
\affiliation{Massachusetts Institute of Technology, Cambridge, MA 02139}
\author{Bryce Meredig}
\email{bryce@citrine.io}
\affiliation{Citrine Informatics, Redwood City, CA 94063}
\author{Venkatasubramanian Viswanathan}
\email{venkvis@cmu.edu}
\affiliation{Carnegie Mellon University, Pittsburgh, PA 15213}


\maketitle


\section{Human Lagtime Model}

When defining our human lag model, we invoke the following assumptions:

\begin{enumerate}
  \item 3 windows of time:
  \begin{enumerate}
    \item Researcher at work (9am-5pm): checks on job every couple of hours;
      average lag of 1 hour.
    \item Researcher (partially) away from work (5pm-11pm): checks on job at
      the end of the window; average lag of 3 hours.
    \item Researcher (completely) away from work (11pm-9am): checks on job at
      the end of window; average lag of 5 hours
  \end{enumerate}

  \item Uniform distribution of when jobs will finish through a week:
  \begin{enumerate}
    \item Probability of job finishing during the week = 5/7: researcher checks
      on job according to 1, above.
    \item Probability of job finishing during the weekend = 2/7: researcher
      checks on job once during the weekend; average lag of 24 hours.
  \end{enumerate}
\end{enumerate}

\section{Surrogate Accuracy}

The size of the unexplored design space shrinks as the simulated sequential
learning (SL) progresses, i.e., the number of candidates in the full dataset
that the model has not ``seen'' yet keeps continuously decreasing.
So surrogate model accuracy estimates derived using model predictions over the
entire unexplored design space (the test set) in each SL iteration can be
affected by the continuously diminishing test set size.
In order to mitigate this effect of test set size on the surrogate model
accuracy estimates, we employ a bootstrapping approach to keep the test set
size fixed in each SL iteration.
At each SL iteration, 20 ``bootstrap test samples'' are generated from the full
unexplored design space.
Each of the 20 bootstrap test samples are generated by randomly sampling, with
replacement, 100 candidates from the unexplored design space.
For each of the 20 bootstrap test samples (with 100 candidates each) we
calculate the mean absolute error (MAE).
Finally, we run 20 independent trials of the entire simulated SL pipeline.
The accuracy of the surrogate model at a given SL iteration is then defined by
the mean and standard deviation of the MAEs of the bootstrap test samples
(generated at that SL iteration) aggregated over the 20 independent trials.
We use this final mean MAE of the surrogate model as the accuracy metric of
interest with a target value of 0.1~eV.

\section{Data and Scripts for Reproducibility}

All data and Python scripts required to perform the analysis presented in this
work are made available via the GitHub repository at
\url{https://github.com/aced-differentiate/closed-loop-acceleration-benchmarks}.

The repository is organized as follows:

\begin{enumerate}
\item
  \href{https://github.com/aced-differentiate/closed-loop-acceleration-benchmarks/data}{data/}

  \begin{itemize}
  \item
    \texttt{benchmark\_calculations\_record.xlsx}: Excel spreadsheet
    containing a record of DFT calculations, associated raw timestamps,
    and a tabulation of the acceleration estimates.
  \item
    \href{https://github.com/aced-differentiate/closed-loop-acceleration-benchmarks/data/bimetallic_catalysts_dataset}{bimetallic\_catalysts\_dataset/}

    \begin{itemize}
    \item
      \texttt{ma\_2015\_bimetallics\_raw.json.gz}: Dataset of bimetallic
      alloys for CO2 reduction, in the
      \href{https://citrineinformatics.github.io/pif-documentation}{Physical
      Information File (PIF)} format, obtained from
      \href{https://citrination.com/datasets/153450}{Dataset 153450} on
      Citrination.

      Original data source: ``Machine-Learning-Augmented Chemisorption
      Model for CO2 Electroreduction Catalyst Screening'', Ma et al.,
      \emph{J. Phys. Chem. Lett.} \textbf{6} 3528-3533 (2015). DOI:
      \href{http://dx.doi.org/10.1021/acs.jpclett.5b01660}{10.1021/acs.jpclett.5b01660}
    \item
      \texttt{transform.py}: Python script for converting from the PIF
      format into tabular data.
    \item
      \texttt{bimetallics\_data.csv}: Bimetallics catalysts dataset
      mentioned above in a tabular format.
    \end{itemize}

  \item
    \href{https://github.com/aced-differentiate/closed-loop-acceleration-benchmarks/data/runtime_geometries}{runtime\_geometries/}

    ``Chemically-informed'' and naive structures and settings in the
    form of \texttt{ase.traj} files, corresponding to the discussion
    surrounding Figure 3 in the paper. The files can be read using ASE
    package (using
    \href{https://wiki.fysik.dtu.dk/ase/ase/io/io.html\#ase.io.read}{\texttt{ase.io.read}}).
  \end{itemize}

\item
  \href{https://github.com/aced-differentiate/closed-loop-acceleration-benchmarks/scripts}{scripts/}

  \begin{itemize}
  \item
    \texttt{human\_lagtime.py}: Script for estimating human lagtime in
    job management, calculated using a Monte Carlo sampling method.
  \item
    \texttt{sequential\_learning.py}: Script for running multiple
    independent trials of sequential learning (SL) and recording a
    history of training examples, model predictions and prediction
    uncertainties.

    If run as-is, the script performs 20 independent trials of 100 SL
    iterations to optimize the \texttt{binding\_energy\_of\_adsorbed}
    property in the bimetallic catalysts dataset mentioned above, using
    three acquisition functions (results from each recorded separately):
    random, maximum likelihood of improvement (MLI) and maximum
    uncertainty (MU).
  \item
    \texttt{plot\_acceleration\_from\_sequential\_learning.py}: Script
    to aggregate results from the \texttt{sequential\_learning.py}
    script, calculate and plot statistics related to acceleration from
    SL over a baseline.

    If run as-is, the script reproduces the 3-paneled Figure 5 in the
    paper.
  \end{itemize}
\end{enumerate}

\subsection{Running the scripts}\label{running-the-scripts}

The required packages for executing the scripts are specified in
\texttt{requirements.txt}, and can be installed in a new environment
(e.g.~using
\href{https://docs.conda.io/projects/conda/en/latest/index.html}{conda})
as follows:

\begin{verbatim}
$ conda create -n accel_benchmarking python=3.10
$ conda activate accel_benchmarking
$ pip install -r requirements.txt
\end{verbatim}

The scripts are all in python, and can be run from the command line. For
example:

\begin{verbatim}
$ cd scripts
$ python sequential_learning.py
\end{verbatim}